\documentclass[11pt]{article}
\usepackage{cospar}
\usepackage{url}

\usepackage{graphicx}

\newcommand{\xmm}{\mbox{\em XMM-Newton }}
\newcommand{\ros}{\mbox{\em ROSAT }}
\newcommand{\asca}{\mbox{\em ASCA }}
\newcommand{\etal}{\mbox{et\ al.\ }}

\newcommand{\ergcms}{\,\mbox{$\mbox{erg}\,~\mbox{cm}^{-2}\,~\mbox{s}^{-1}$}}

\newcommand{\msun}{\,\mbox{$\mbox{M}_{\odot}$}}

\newcommand{\aanda}  {{\em Astron. \& Astrophys.}\nolinebreak }
\newcommand{\apj}    {{\em Astrophys. J.}\nolinebreak}

\newcommand{\mn}     {{\em Mon. Not. Royal Ast. Soc.}\nolinebreak }

\title{THE OPTICAL TO X-RAY SPECTRUM OF THE BROAD-LINE ULTRASOFT AGN RE J2248-511}

\author{R. L. C. Starling\address{Mullard Space Science Laboratory, University College London, Holmbury
St. Mary, Dorking, Surrey RH5 6NT, UK},
        E. M. Puchnarewicz$^{1}$,
        E. Romero-Colmenero\address{South African Astronomical Observatory, PO Box 9, Observatory, 7395 South
Africa} 
	and K. O. Mason$^{1}$}

\begin{document}

\maketitle

\begin{abstract}

We present a series of monitoring observations of the
ultrasoft broad-line Seyfert galaxy RE J2248-511 with \xmm. Previous
X-ray observations showed a transition from a very soft state to a harder
state five years later. We find that the ultrasoft X-ray excess has re-emerged, yet there is no change in the hard power-law. Reflection models with a
reflection fraction of $\ge$ 15, and Comptonisation models
with two components of different temperatures and optical depths (kT$_{1}$ = 83
keV, T$_{1}$ = 30 eV, $\tau_{1}$ = 0.8; kT$_{2}$ = 3.5
keV, T$_{2}$ = 60 eV, $\tau_{2}$ = 2.8) can be fit to
the spectrum, but cannot be constrained. The best representation of the
spectrum is a model consisting of two blackbodies (kT$_{1}=0.09\pm0.01$ keV,
kT$_{2}=0.21\pm0.03$ keV) plus a power-law ($\Gamma=1.8\pm0.08$). 
We also present simultaneous optical and infrared data showing that the
optical spectral slope also changes dramatically on timescales of
years. If the optical to X-ray flux comes primarily from a Comptonised
accretion disk we obtain estimates for the black hole mass 
$\sim$10$^{8} \msun$, accretion rate $\sim$0.8$\dot{M}_{Edd}$ and inclination
$\cos(i) \ge 0.8$ of the disk. 
\end{abstract}

\section*{INTRODUCTION}

Studies of \ros PSPC slopes in AGN have concluded that sources with
steep soft X-ray (0.1-2.4 keV) continuum slopes in addition to broad
optical emission lines (H$\beta$ FWHM$\ge$3000 km s$^{-1}$) are simply
not found in nature (Boller, Brandt \& Fink 1996; Wandel \& Boller
1998). The steep soft X-ray slopes characteristic of Narrow-line
Seyfert 1 galaxies (NLS1s; whose H$\beta$ FWHM$\le$2000 km s$^{-1}$)
are thought to be due to a small mass black hole with a high accretion
rate (eg Pounds, Done \& Osborne 1995) or perhaps due to a broad line
region (BLR) which forms relatively far out from the central black
hole (Puchnarewicz et al. 1992).

RE J2248-511 is a radio-quiet EUV-selected AGN ($z$ = 0.101) discovered
by the \ros Wide Field Camera (Pounds et al. 1993), and classified as
a normal broad-line Seyfert 1 galaxy (Mason et al. 1995) with
H$\beta$ FWHM$\sim$3000 km s$^{-1}$.  A subsequent observation with
the \ros PSPC (Puchnarewicz et al. 1995) found that the soft X-ray
spectrum was very soft and was best-fitted with a broken power law where
$\Gamma$ = 4.13 at energies below 0.25 keV (and $\Gamma$ = 2.62 at
higher energies). Thus it differs from a NLS1 in that it has {\sl
high} velocity BLR clouds, yet a very steep, ultrasoft X-ray
spectrum. This puts it in a category comprising, to our knowledge, only 3 AGN
with both these characteristics (RE~J2248-511, 1H~0419-577 and RXJ0437.4-4711).

RE~J2248-511 was also observed by \asca in 1997 (Breeveld et al. 2001). These
observations showed that the soft excess (below 2.0 keV) had
completely disappeared and the entire 0.5-10 keV spectrum could be
represented by a single power law. Large amplitude variability in the
soft X-ray region has also been observed in another of the broad-line
ultrasoft AGN 1H~0419-577 (Guainazzi et al. 1998, Turner et al. 1999,
Page et al. 2002). This two-state behaviour has led to comparisons
with Galactic black holes (GBHs). Page \etal (2001) show
that the 0.2-10 keV spectrum of 1H~0419-577 closely resembles that of
two Comptonised components, although these cannot be well constrained,
and that the variability between \xmm and \asca is explained by a
change in optical depth of the Comptonising material. Comptonisation
has also been proposed as a possible source for the X-ray emission of
RE J2248-511 (Breeveld et al. 2001) which we will investigate in this paper.

Here, we present the results of a recent multi-wavelength monitoring
campaign consisting of observations from \xmm and the South
African Astronomical Observatory (SAAO), to examine the long-term
variability of this source and the relationship between the X-ray
emitting components and the response of the BLR.

\section*{XMM-NEWTON OBSERVATIONS}

RE J2248-511 was observed by the EPIC (Str\"uder et al. 2001; Turner et al. 2001) and RGS
(den Herder et al. 2001) instruments on-board \xmm (Jansen et
al. 2001) on 2000 October 26 (17.6 ks pn), 2001 April 21
($\sim$8.5 ks) and 2001 October 31 (15.4 ks pn). The
Optical Monitor was switched off since
a nearby bright star saturates the field. All EPIC instruments were in
small window mode through the medium filter. About
4.5 ks of the 2001 April observation was contaminated by an extremely
high background flare which makes the source difficult to distinguish
above the noise. For this reason we do not use these data here. We
will concentrate on the EPIC pn data for this analysis since the MOS cameras were in small window free running
mode which has not yet been fully calibrated, and the pn data
have the higher net count rate.

The raw data from all observations were processed with the \xmm SAS
(Science Analysis Software) version 5.3. We extracted a circular
region around the source with radius 41.3 arcsec. Background
subtraction was done using a nearby source-free region
4x larger in radius than the source extraction region. We 
summed the spectra from the 2000 October and the 2001 October pn
observations to increase signal to noise, having first confirmed that no
significant variability occurs either during or between the two
observations. The latest EPIC pn response file for small window
medium filter single events has been used. The resulting spectrum is
grouped such that a minimum of 20 counts fall in each bin so
$\chi^{2}$ statistics apply. The mean net count rate is 5.71$\pm$0.02 counts s$^{-1}$.

\section*{X-RAY SPECTRAL ANALYSIS}

Spectral analysis was done using the X-ray spectral fitting package
XSPEC, using the $\chi^{2}$ minimization technique. We have restricted
the considered energy range to 0.3-10.0 keV as there may be
uncertainties in the instrument calibration at energies outside this
range. We find that a simple power-law cannot fit the spectrum over
this range largely due to the presence of a strong soft X-ray excess (Figure 1).
\begin{minipage}{70mm}
\includegraphics[width=60mm, angle=-90]{figure1.ps}

{\sf Fig. 1. The summed EPIC pn spectrum is shown against a power-law model with galactic absorption.}
\end{minipage}
\hfil\hspace{\fill}
\begin{minipage}{70mm}
\includegraphics[width=60mm, angle=-90]{figure2.ps}

{\sf Fig. 2. The unfolded EPIC pn spectrum with the best-fitting model of 2
blackbodies and a power-law plus galactic absorption. }
\end{minipage}

The pn spectrum is well fitted with a model consisting of two blackbodies
dominating at low energies with temperatures kT$_{1}=0.09\pm0.01$ keV and
kT$_{2}=0.21\pm0.03$ keV, plus a power-law with $\Gamma$ =
1.8$\pm$0.08 ($\chi^2$/dof = 458/468; Figure 2). The power-law slope is consistent with the
overall 2-10~keV spectrum of Seyferts in general, which have indices
typically $\Gamma\sim$1.8-2.0. The X-ray flux in the 0.3-10.0 keV
energy range is 9.83$\times$10$^{-12}$ \ergcms, corresponding to a
luminosity of 4.74 $\times$10$^{44}$ erg s$^{-1}$ (assuming H$_{0}$ =
50.0 and q$_{0}$ = 0.0). Upper limits on the equivalent widths of an
Iron line at 6.4 keV (in the rest frame of the source) are 60 eV for
a narrow line ($\sigma$ = 0.01) and 50 eV for a broad line ($\sigma$
= 0.3). Inclusion of the narrow line gives an improvement according
to the F-test of F statistic value = 2.14 and probability = 9.7$\times$10$^{-2}$.
The Galactic column density was
fixed at 1.4$\times$10$^{20}$ cm$^{-2}$ (Puchnarewicz et
al. 1995) in all fits. Adding an intrinsic absorption
component gives no significant improvement to the fit.

\subsection*{Physical Models}
We tried a simple power-law-with-reflection model (Magdziarz \& Zdziarski
1995), where incoming photons may be reflected
off ionised material such as an accretion disk. This gives a good fit only if
the X-ray flux is reflected with a reflection fraction of $\ge$15
and $\chi^{2}$/dof = 721/659.

Comptonisation has often been suggested as a source of both the soft X-ray and
hard X-ray spectrum so we investigated a number of such possibilties using the
comptt model within XSPEC (Titarchuk 1994). The best fitting model (
$\chi^{2}$/dof = 672/656; Figure 3) is a hot optically thin component
(kT$_{1}$ = 83
keV, $\tau_{1}$ = 0.8) irradiated by a 30 eV photon
temperature, alongside a cooler optically thick component (kT$_{2}$ = 3.5
keV, $\tau_{2}$ = 2.8) irradiated by a 60 eV
temperature medium, both upscattering the incident photons. However, while the parameters
may appear physically realistic they cannot be reasonably constrained. 

We also note that a multi-temperature disk blackbody plus Comptonisation can
also fit the spectrum well with $\chi^{2}$/dof = 713/654, but again the
parameters are unconstrained.

\begin{minipage}{70mm}
\includegraphics[width=60mm, angle=-90]{figure3.ps}

{\sf Fig. 3. The unfolded EPIC pn spectrum and double Comptonisation model.}
\end{minipage}
\hfil\hspace{\fill}
\begin{minipage}{70mm}
\includegraphics[width=60mm, angle=-90]{figure4.ps}

{\sf Fig. 4. The \xmm EPIC pn data in the 0.7-10~keV range,
together with the power-law model which best fits the \asca GIS2 and GIS3 data. The lower
panel shows the ratio of the fit to the data, clearly showing excess emission in the pn continuum below 1~keV.}
\end{minipage}

\section*{COMPARISON WITH \asca AND \ros} 

Comparing the data from all three missions (see also Puchnarewicz et
al. 1995 and Breeveld et al. 2001) in the overlapping energy range of
0.7-2.0 keV by simply fitting a power-law (Table 1), we find that in
the \xmm observations RE J2248-511 is in a much softer state than at
the time of the \asca observation ($\Gamma$ = 2.46 in \xmm but 2.05 in
\asca; for \asca, the SIS response is uncertain below 1~keV, so only
the GIS data were used). The power-law slope measured with the \ros
PSPC is softer still ($\Gamma$ = 2.7), although it should be borne in
mind that there are reports that the PSPC may over-estimate $\Gamma$
by as much as 0.4 in which case the \xmm and \ros PSPC data would
not be inconsistent. We also note that the error on the PSPC slope is large enough to be consistent with either the \xmm
or \asca slopes over this range.

Thus, the soft X-ray component first observed with \ros which
subsequently disappeared in the \asca observation (Breeveld et
al. 2001), now seems to have re-appeared in \xmm (Figure 4). The 2-10~keV
spectrum has changed relatively little (in flux and spectral form)
between the \asca and \xmm observations. From the
two-blackbody-plus-power-law fit to the \xmm EPIC pn data (kT$_{1}=0.09\pm0.01$ keV,
kT$_{2}=0.21\pm0.03$ keV, $\Gamma=1.8\pm0.08$), we calculate a flux in the soft
component (ie in the two blackbodies; 0.3-2.0 keV) of
2.49$\times$10$^{-12}$ \ergcms.

\begin{table}[t]
\caption{Power-law fits to the \ros PSPC, \asca GIS 2 and 3 (fitted
simultaneously) and \xmm EPIC pn data in the overlapping range of 0.7-2.0 keV;
z = 0.101, Galactic absorption = 1.4$\times$10$^{20}$ cm$^{-2}$ (both fixed), no intrinsic absorption.}
\begin{tabular}{lccc}
\hline
\\
 &$\Gamma$ & norm & $\chi^{2}$/dof\\
 & & ($\times$10$^{-3}$) & \\
\ros PSPC (1993) &  2.72$^{+0.98}_{-1.38}$ &3.67$\pm$0.131& 25.3/17 \\
\asca GIS (1997)& 2.05$\pm$0.11& 1.99$\pm$0.175  &  60.3/58 \\ 
\xmm EPIC pn (2000-01)& 2.46$\pm$0.03 &2.25$\pm$0.037&   235/240 \\ 
\hline
\end{tabular}
\end{table}

\section*{OPTICAL AND IR OBSERVATIONS}

RE J2248-511 was observed on 2000 October 19-24 using the South
African Astronomical Observatory's 1.9m Radcliffe telescope. Three
observations were made with the grating spectrograph and SITe CCD one
night apart and measurements of the IR magnitude in the J,H,K bands
were made on a single night with the infrared photometer. Spectra were taken using both a wide (700 $\mu$m) and a narrow (300 $\mu$m) slit with overlapping red and blue gratings
($\lambda_{central}$ = 4600,7800\AA). Arc spectra were taken before and
after every target and every standard spectrum. The spectra have been
extracted using the Image Reduction and Analysis Facility (IRAF). 

\section*{RESULTS FROM THE OPTICAL SPECTROPHOTOMETRY}

Seeing, humidity and cloud cover were extremely variable during the observing
period and consequently only one of the nights we consider to be photometric.
The spectra were not taken at the parallactic angle, but there appears to have
been little or no loss of blue light since the spectra become stronger towards
the blue end. 

\subsection*{The Broad Line Region}
Both H$\alpha$ and H$\beta$ are best fitted with three gaussians with broad
components of FWHM approximately 3700\AA. The ratio of H$\alpha$ to
H$\beta$ line flux, excluding the narrow-line component, is H$\alpha$/H$\beta$ = 2.67.
The line parameters we measure are consistent with those measured in the
spectrum during 1991 (Mason et al. 1995; H$\alpha$ was not observed). The
H$\beta$ measurements are in approximate agreement with the 1992 measured
values (Grupe et al. 1999). 

\begin{minipage}{70mm}
\includegraphics[width=60mm, angle=90]{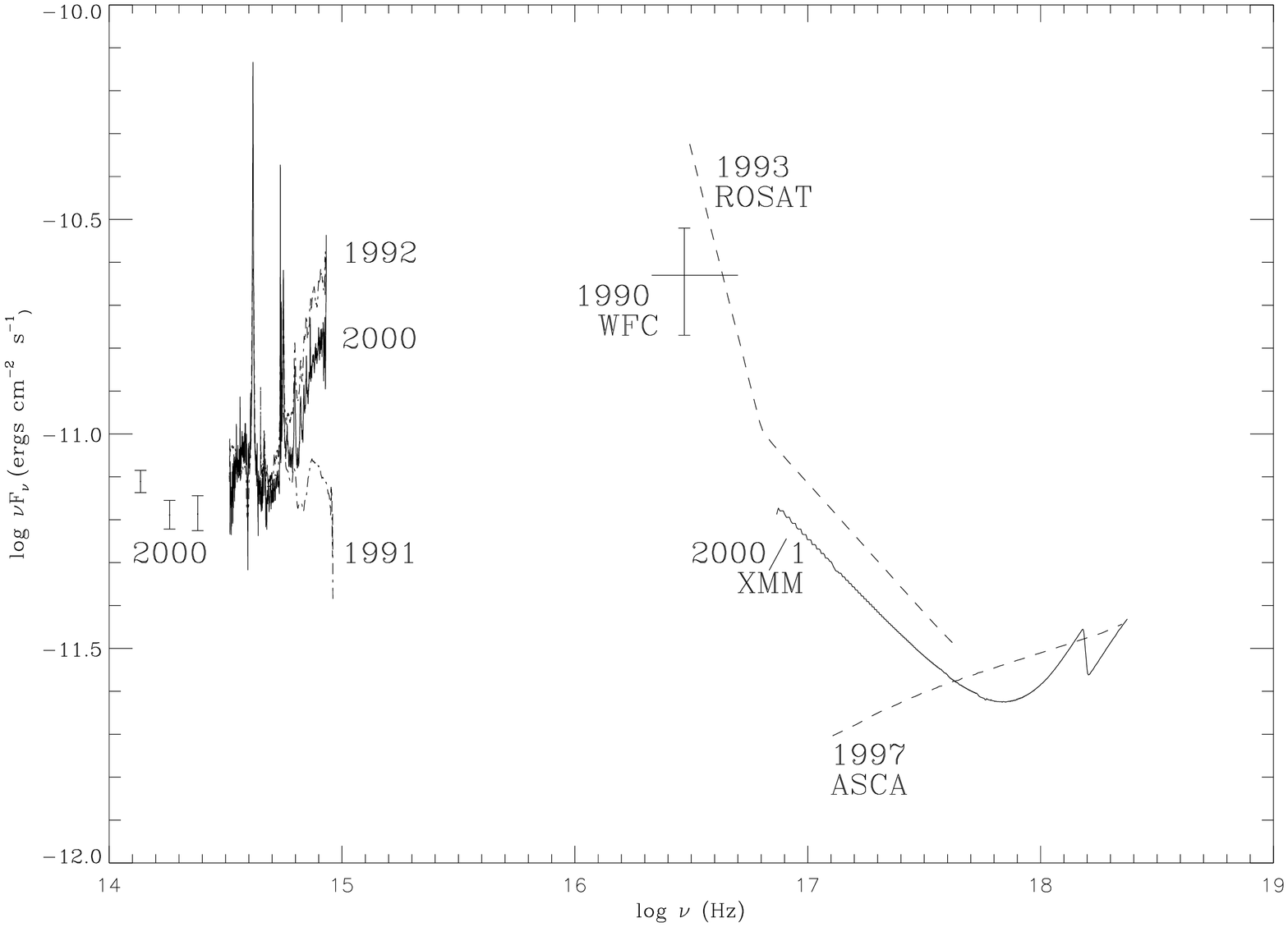}
{\sf Fig. 5. Spectral energy distribution from optical to X-ray for RE
J2248-511 including all available optical, IR and UV data and best-fitting X-ray models spanning 1990 to 2001.}
\end{minipage}
\hfil\hspace{\fill}
\begin{minipage}{70mm}
\includegraphics[width=60mm, angle=90]{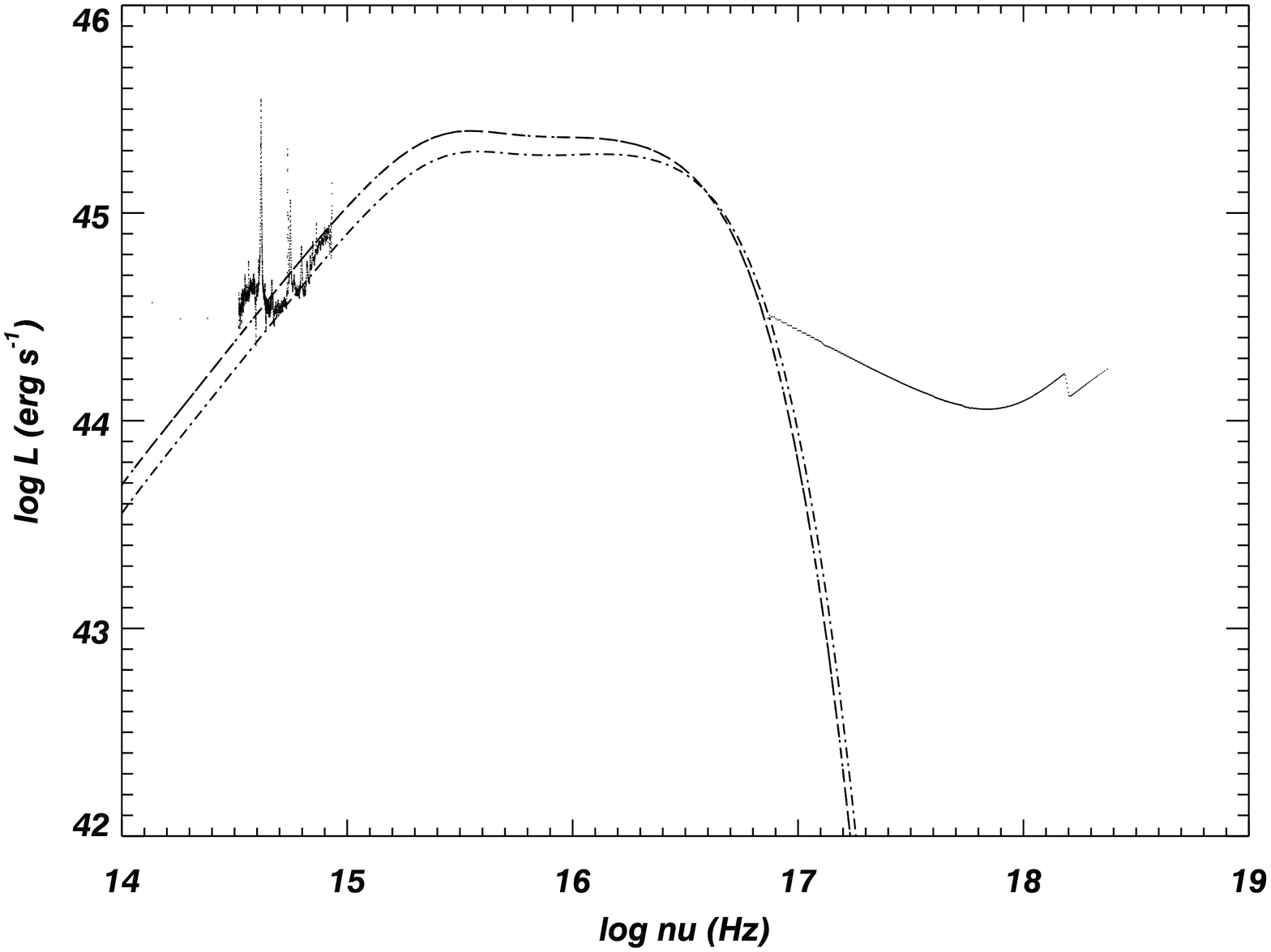}
{\sf Fig. 6. The
SED for October 2000 and two Comptonised disk models with $\dot{M}=0.8M_{Edd}$ and $\cos(i)=0.8$. The model represented
by the dashed line is for black hole mass $M=10^8\msun$; the dot-dashed line
is for $M=10^{7.9}\msun$.}
\end{minipage}

\subsection*{Optical Variability and the Big Blue Bump}   
In Figure 5 all available data for RE J2248-511 is plotted to show the various flux levels that
have been observed. We find that the slope of the optical spectrum in the SED plot lies in between the
slopes measured in 1991 (Mason et al. 1995) and 1992 (Grupe et al. 1998).
Since the soft X-ray data from 2000/01 can be fit with a multi-temperature disk blackbody
model, we
investigated the possibility of the entire optical to soft X-ray flux
originating in an accretion disk. We use Comptonised accretion disk models
developed by Czerny \& Elvis (1987) and fitted these to the spectral energy
distribution for October 2000. Two model fits are shown with the data in
Figure 6. These models can fit the optical slope reasonably well and are
not inconsistent with the soft X-ray data. These data require models with
masses of order $M=10^{8}\msun$, high accretion rates of
$\dot{M}\sim0.8M_{Edd}$ and high disk inclination angles of $\cos(i)\ge0.8$. 

\section*{DISCUSSION}

Of the physical models for the X-ray spectrum tested here we are unable to constrain the key
parameters in any case.
The power-law-plus-reflection model is able to fit the data only with a
high reflection fraction. This multiple reflection scenario could conceivably
occur if there were a large number of clouds above or around the X-ray source
which could reflect the emission but not transmit it. Alternatively, the
accretion disk could be of varying height, like ripples on a pond, each ripple
reflecting the incoming X-rays (Ballantyne, Ross \& Fabian 2001). However, we
do not detect strong Iron emission or the neutral Iron edge at 7.1 keV which
are characteristic of strong reflection, except where the reflecting material is
highly ionised. 
This Seyfert lies at the boundary between
the quasar and Seyfert galaxy luminosities. Seyfert galaxy X-ray spectra typically show disk features
including Iron lines and a reflection component which are less common in
quasars. This is thought to be because the disk material becomes fully ionised
and does not produce emission lines (eg. Matt et al. 1993). Though we do not
see low ionisation Iron K$\alpha$ emission in these data, the upper limits on its
equivalent width are typical of lines already observed in many Seyfert galaxies. Therefore reflection is unlikely to be making a strong
contribution to the X-ray continuum flux in this source.

The double Comptonisation model requires both optically thick and optically
thin material, each being illuminated by different temperature photons.   
A change
in optical depth of the material could explain the change between the flat \asca
spectrum and the very soft EPIC pn spectrum. This scenario has also been
suggested for the Seyfert 1 galaxy 1H 0419-577 (Page et al. 2001) and is
discussed for this source in Breeveld et al. (2001). This model, however, could not account for any correlation with the optical
flux.

Currently there is insufficient data to distinguish between many of the
models and a 2-blackbodies-plus-power-law model remains
the best representation of this data. \\
From the variability time-scale we estimate the size of the region emitting the
flux of the soft X-ray excess to be $\le$3.4 l.y. or $\le$1 pc, which only locates it to at or within the BLR.  

That both the 2000 optical and soft X-ray slopes 
are steep and both wavebands show great variability suggests a link between them. The optical to soft X-ray spectral energy distribution for 2000 is not inconsistent
with Comptonised accretion disk models. Contrary to the proposed NLS1 scenario, the
data are best fitted with intermediate to high black hole masses, whilst still
favouring the high accretion rates suggested for NLS1 galaxies. Therefore, it
is not necessary for a black hole to have a low mass for the formation of an
ultrasoft X-ray excess. The relatively high velocity of the BLR may, however,
be linked with the black hole mass, perhaps as a consequence of the increased
gravitational potential. The mass and accretion rate we estimate from this
modelling implies an inner disk effective temperature of $\sim$25 eV assuming a
standard disk model, lower
than that fitted to the lower-temperature soft X-ray component. However, the
temperature measured in the X-ray blackbody models is the colour temperature
which may be much higher than the effective temperature if the spectrum is
modified by electron scattering (eg Titarchuk \& Shrader 2002). 

\section*{CONCLUSIONS}

This \xmm observation has shown yet more dramatic activity in the soft X-ray
spectrum of RE J2248-511, whilst the hard X-ray power law remains constant. 
We find
that both Comptonisation and reflection models, while appearing to fit the
X-ray spectrum well, cannot be
constrained, and a model consisting of 2 blackbodies and a power-law remains
the best representation of these data. The 2000 optical to soft X-ray
spectral energy distribution may be consistent with a face-on Comptonised
accretion disk accreting at a rate approaching the Eddington limit onto a 10$^8 \msun$ black
hole. Observations at UV wavelengths are needed to tie down the parameters of any such
model, and more simultaneous multi-wavelength data are needed to confirm the
correlation between optical and soft X-ray variability and the presence of a
single Big Blue Bump.   

\section*{ACKNOWLEDGEMENTS}

This work is based on observations obtained with \xmm, an ESA science
 mission with instruments and contributions directly funded by ESA Member
 States and the USA (NASA).
IRAF is distributed by the National
 Optical Astronomy Observatories, which are operated by AURA, Inc., under
 cooperative agreement with the National Science Foundation. We would like to thank Aneta Siemiginowska for use
 of the Comptonised accretion disk models. RLCS acknowledges support from a PPARC studentship.

email address of R. L. C. Starling    rlcs@mssl.ucl.ac.uk \\
Manuscript received 10/12/02; revised 11/02/03; accepted 13/02/03
\end{document}